\documentclass{ws-procs975x65}            
\pdfoutput=1
\usepackage{amsmath,amssymb,amsbsy,booktabs,array,rotating,verbatim}
\usepackage{bm,comment}
\usepackage{graphicx,color,ulem}
\usepackage[sort&compress]{natbib}
\usepackage[colorlinks=true, linkcolor=blue, filecolor=blue, urlcolor=blue, citecolor=blue, pdftex, plainpages=false]{hyperref}

\newcommand\buaff{Department of Physics and Center for Computational Science, Boston University,\\ Boston, MA 02215, USA}
\newcommand\coaff{Department of Physics, University of Colorado,\\ Boulder, CO 80309, USA}
\newcommand\Higgsaff{Higgs Centre for Theoretical Physics, School of Physics \& Astronomy,\\ The University of Edinburgh, Edinburgh, EH9 3FD, UK}

\newcommand{\gGF}{\ensuremath{g_{\rm GF}^2} }

\newcommand{\gtGF}{\ensuremath{\widetilde g_{\rm GF}^2} }

\definecolor{orange}{rgb}{1.0, 0.5, 0}
\definecolor{Green}{rgb}{0, 0.588, 0}

\begin{document}
\title{Strongly coupled gauge theories:\\ What can lattice calculations teach us?}

\author{A. Hasenfratz$^*$ }

\address{\coaff\\
$^*$E-mail: Anna.Hasenfratz@colorado.edu}

\author{R.C.~Brower, C.~Rebbi and E.~Weinberg}

\address{\buaff}

\author{O.~Witzel}
\address{\Higgsaff}

\begin{abstract}
The dynamical origin of electroweak symmetry breaking is an open
question with many possible theoretical explanations. Strongly coupled systems predicting the Higgs boson as a bound state of a new gauge-fermion interaction form one class of candidate models. Due to increased statistics, LHC run II will further constrain the phenomenologically viable models in the near future. In the meanwhile it is important to understand the general properties and specific features of the different competing models.

In this work we discuss many-flavor gauge-fermion systems that contain both massless (light) and massive fermions. The former provide Goldstone bosons and trigger electroweak symmetry breaking, while the latter indirectly influence the infrared dynamics.
Numerical results reveal that such  systems can exhibit a light
$0^{++}$ isosinglet scalar, well separated from the rest of the spectrum. Further, when we set the scale via the $vev$ of electroweak symmetry breaking, we predict a 2 TeV vector resonance which could be a
generic feature of SU(3) gauge theories.
\end{abstract}


\bodymatter

\section{Introduction}

The  Standard Model  is a very successful description of electromagnetic, weak and strong interactions, yet the dynamical nature of its central feature, electroweak symmetry breaking (EWSB), has remained elusive.  While experimental data  strongly constrains physics Beyond the Standard Model (BSM), many options are still viable. New LHC results,  such as the recently published  di-boson resonance around 2 TeV, hint at new interactions \cite{Aad:2015owa, CMS:2015gla}. Increased statistics of  LHC run II might finally reveal  the origin of the Higgs boson and give insight into BSM physics.

Strongly coupled  gauge-fermion systems are among the major  contenders to describe BSM dynamics. In these models a new   strongly interacting sector gives rise to a  new set of hadronic states that lead to experimentally verifiable predictions. While weakly coupled ``QCD-like'' systems are not compatible with electroweak precision measurements, recent non-perturbative lattice calculations show that strongly coupled systems, especially   those near the conformal window, have properties that are very different from the weakly coupled models,  making them good BSM candidates.  
There are several promising systems with different gauge and fermion content that have received significant attention lately, see e.g.~\cite{Aad:2014cka,Arbey:2015exa,Aoki:2014oha,Hietanen:2014xca,Fodor:2015vwa,Schaich:2015psa}. In this paper we describe  a simple model that, while   maybe not phenomenologically viable,  can serve as  basis of future, more sophisticated, and phenomenologically better motivated investigations. Preliminary results of our study have been presented in \cite{Brower:2014dfa,Brower:2014ita}.

The strongly coupled BSM candidate models we are interested in must exhibit spontaneous  chiral symmetry breaking. Their infrared (IR) spectrum   contains  three massless Goldstone pions and  a relatively light $0^{++}$ scalar, in addition to a tower of additional massive states. When coupled to the electroweak  sector the Goldstone pions become the longitudinal components of the $W$ and $Z$ bosons, triggering electroweak symmetry breaking.  In this scenario the pseudoscalar (pion) decay constant $F_\pi$ sets the scale of the BSM model, $F_\pi \approx 250$ GeV, the $vev$ of  EWSB.  Among  the massive states the $0^{++}$ meson plays the role of the Higgs boson, while the other states predicted by the model appear as heavy resonances at higher energies. Even though radiative corrections can significantly alter the mass of the $0^{++}$ scalar, for phenomenological applications it is desirable to have the isosinglet scalar well separated from the higher excitations. 

 If the model has only two massless fermions, there are only three Goldstone pions in the  IR. However many models rely on more fermion flavors to push the system closer to the  conformal window where non-perturbative effects  can change  the  IR dynamics. 
  In this case the IR spectrum contains many more Goldstone pion states than are required for EWSB. Those massless states have to acquire mass via a sequence of symmetry breaking steps between the UV and IR energy regions.  While the explicit mechanism   is model dependent and can be complicated,  we can approximate the steps by lifting  the mass of all but two fermions with explicit mass terms   allowing them to  decouple in the IR limit.
 
Theories that are either chirally broken or conformal  but close to the conformal window  can  look very similar in a finite volume lattice study.  Nevertheless, it is essential to ensure that the BSM candidate model breaks chiral symmetry spontaneously. If a gauge-fermion system turns out  to be  conformal, it has to be driven  below the conformal window. This can easily be done by lifting the fermion masses in many-fermion systems as we do in this work.  Alternatively,  other interactions, possibly 4-fermion terms, may be added to ensure spontaneous chiral symmetry breaking. 

\section{ The 4+8 flavor model}

 The model we consider in this work is based on the SU(3) gauge theory with $N_f=12$ fundamental flavors. This system is conformal when all 12 flavors are massless \cite{Cheng:2013eu,Cheng:2013xha,Lombardo:2014pda,Itou:2013ofa,Itou:2014ota}.  Ideally we would like to lift the mass of 10 of the flavors to end up with a system that contains only 3 Goldstone pions in the IR. In this pilot study, however, we give mass only to 8 fermion species, keeping 4 in the chiral limit. The reason  for this choice is technical. We use staggered fermions that naturally come in multiplets of four.  It is possible to split the masses of the multiplets by taking the fourth root of the fermion determinant but taking the continuum limit requires special care, especially in strongly coupled systems. To avoid any potentially uncontrolled lattice artifacts we decided to consider a system with 4 massless (or at least light) and 8 heavy flavors.  
 
  Even if we kept only two fermions light,  the $N_f=12$ model is not ideal as it has a relatively small mass anomalous dimension. As we will argue in the next section, the anomalous dimension dictates scaling in  the mass-split system as well.  Systems closer to the conformal boundary, like $N_f=10$ or perhaps 8 flavors could lead to phenomenologically more interesting systems. We are considering  to investigate systems with 2+6 or 2+8 flavors in the future.
 
\section{Wilson renormalization group description}
\label{sec:rgequations}
 
The Wilson renormalization group (RG) approach applied for conformal and mass deformed conformal systems predicts the general scaling behavior of all dimensional quantities. In the following brief review  we follow the notation of Ref.~\cite{DelDebbio:2010ze}, but  extend that approach  by allowing only a subset of the  fermions to become massive while keeping the rest in the massless chiral limit.

Consider a 4-dimensional gauge-fermion system characterized by a set of gauge couplings $g_i$ and masses  $m_i$. Perturbatively, the  gauge couplings are dimensionless while the dimension of the fermion mass is carried by the lattice cutoff. For convenience we introduce 
\begin{align}
\widehat m_i = a m_i \propto  m_i / \Lambda_a,
\end{align}
with  $\widehat m_i $  the dimensionless ``lattice mass'' and $\Lambda_a = \pi/a$  the lattice cutoff.

At the perturbative UVFP one of the gauge couplings, $g_1$, is marginally relevant (the usual asymptotically free gauge coupling), while all other gauge couplings are irrelevant.  The $\widehat m_i$ mass couplings are relevant at the perturbative UVFP and we assume they will stay relevant everywhere in the parameter space considered. 

We choose the number of flavors such that the system is still asymptotically free but  inside the conformal window, i.e.~it has an IRFP at ($g_i=g_i^\star$, $\widehat m_i=0$) where  all of the gauge couplings are irrelevant. If we tune the system  to the conformal FP, it will be conformal at any energy range.  Away but still close to the IRFP in the gauge couplings, at ($g_i\approx g_i^\star$ , $\widehat m_i=0$), there are corrections to conformal behavior. In the IR those die out and  the  system is conformal  everywhere.
Finite mass breaks conformality. If $\widehat m_i \gtrsim 0$ we  have a mass-deformed system. 

For the remainder of this section we will assume that the  system is in the vicinity of the IRFP, i.e. the bare  parameters as defined at the cutoff scale $\Lambda_a$ are ($g_i \approx g_i^\star$, $\widehat m_i \gtrsim 0$) and we can use Wilsonian RG equations to describe the parameter and energy scale dependence of any 2-point correlation function 
\begin{align}
C_H(t; g_i,\widehat m_i,\mu= \Lambda_a) = \int d^3x \langle H(t,x) H(0)^{\dagger} \rangle \Big |_{g,\widehat m,\Lambda_a}\,.
 \end{align}
 
 In the vicinity of the IRFP an RG transformation that changes the scale 
 \begin{align}
 \mu \to \mu^\prime = \mu /b, \quad  b>1
 \end{align}
 transforms the parameters as 
 \begin{align}
  (g_i(\mu)-g_i^\star) &\to  ( g_i(\mu^\prime)-g_i^\star)   = b^{y_{g_i}} (g_i(\mu)-g_i^\star) , \label{eq:rg_running}\\
   \widehat  m_i(\mu)  & \to  \widehat m_i(\mu^\prime)  = b^{y_{m_i}} \widehat m_i(\mu) \nonumber
 \end{align}
 while the correlation function changes as
\begin{align}
C_H(t; g_i,\widehat m_i,\mu) =  b^{-2\gamma_H} C_H(t; g_i^\prime,\widehat m_i^\prime,\mu^\prime).
\label{eq:C_H1}
 \end{align} 
 Here $\gamma_i$ denotes the anomalous dimension while $y_i = d_i + \gamma_i$ is the scaling dimension of the corresponding operator at the IRFP. 
 Finally we rescale all dimensional quantities in  \eref{eq:C_H1} by $b$  and obtain
 \begin{align}
C_H(t; g_i, \widehat m_i,\mu) =  b^{-2y _H} C_H(t/b; g_i^\prime, \widehat m_i^\prime,\mu),
\label{eq:C_H2}
 \end{align} 
 where for simplicity we do not indicate   the explicit $\mu$ dependence of the couplings.

For the irrelevant couplings the scaling dimensions $y_{i}<0$, while  for  relevant couplings  $y_{i}>0$.   With repeated RG steps $b$ increases and the irrelevant couplings  $g_i$ approach $g_i^\star$. The fermion mass parameters, on the other hand, increase as $\widehat m_i \to  b^{y_{mi}} \widehat m_i$.  The fermions decouple from the IR dynamics when $\widehat m_i^\prime = \mathcal{O}(1)$ because the masses are now above the cutoff. For concreteness we choose $b = \widehat m_i ^{-1/y_m}$, so  $\widehat m(b^{y_m} \mu) =1$ is the scale where the massive fermions no longer affect the IR dynamics.
 
 It is worth noting that we can  define $b$ uniquely  only if all massive fermions start with the same bare mass, $\widehat m_i(\Lambda_a) = \widehat m$. The RG equations become considerably more complicated  if the fermions have different bare mass values.
There is no complication however  if we keep some fermions  massless. These fermions will remain massless and  present in the IR theory. 
 
If $b$ is large enough we can neglect any dependence on $g^\prime -g_\star$ and the RG equation of ~\eref{eq:C_H2} reduces to
 \begin{align}
 C_H(t; \widehat m_i(\mu),\mu) = \mathcal{C}_H F(t (\widehat m)^{1/y_m},\mu),
\label{eq:C_H3}
 \end{align}
 where $F$ is some function that, for fixed $\mu$, depends only on the rescaled variable $t (\widehat m)^{1/y_m}$.
 Any 2-point correlation function is expected to show exponential behavior at large distances,
  \begin{align}
 C_H(t; \widehat m_i,\mu) \propto e^{-M_H t}, \quad \quad t \to \infty.
 \label{eq:C_asymp} 
 \end{align}
 Combining equations (\ref{eq:C_H3}) and (\ref{eq:C_asymp}) leads to the well-known (hyper)scaling relation
\begin{align}
a M_H \propto (\widehat m)^{1/y_m}.
\label{eq:M_scaling}
\end{align}

 If all fermions are massive, Eqs.~(\ref{eq:C_H3}) and (\ref{eq:M_scaling}) describe the scaling behavior of correlation functions of degenerate fermions.  Equations (\ref{eq:C_H3}) and (\ref{eq:M_scaling}) also describe correlation functions of pure gauge quantities and the scaling of glueballs, the square root of the string tension, etc.  In the deep infrared limit the system is an SU($N_c$) gauge theory, no  dynamical fermions are present. However this system is not identical to what we usually refer to as pure-gauge or quenched SU($N_c$) theory. The  fermions, while decoupled in the IR, do influence the IR dynamics. As long as the system evolves from the scaling regime of the IRFP, the scaling relations of Eq.~(\ref{eq:M_scaling}) hold for any bound states.  The dimensionless ratios of physical quantities are expected to be independent of $\widehat m$, up to corrections to scaling  due to both the irrelevant and relevant couplings. 
   
 The situation becomes  more  complicated if we keep some of the fermions massless, i.e.~we have $N_l$ flavors in the $\widehat m_\ell=0$ chiral limit, while we give mass to $N_h$ fermion flavors, $\widehat m_h >0$.  Equations (\ref{eq:C_H3}) and (\ref{eq:M_scaling}) are still valid, but the correlation function $C_H(t)$ can contain  operators  constructed from massless fermions, massive fermions, or a combination of both. If the number of massless fermions, $N_\ell$, is below the conformal window, the system is chirally broken and contains $N_\ell^2-1$ massless Goldstone pions in the IR. It also contains light hadronic states, similar to an $N_l$ flavor SU($N_c$) theory in the chiral limit. States that contain at least one heavy flavor decouple and are not present in the deep IR limit, nevertheless their influence on the light hadron spectrum remains.  All massive states, even those that contain only  massless fermions, show hyperscaling with respect to the heavy flavor mass $\widehat m_h$. In particular, ratios of masses are expected to be independent of $m_h$ (up to corrections due to scaling violations). While these  ratios are governed by the conformal IRFP, their value can depend on $N_\ell$, in addition to $N_\ell + N_h$ that determines the scaling dimensions at the IRFP.

\subsection{The running gauge coupling}

We consider  first the running  coupling of conformal and mass-deformed systems based on the  Wilsonian RG discussion of the preceding section. 

\begin{figure}[thb]
\centering
\parbox{0.49\textwidth}{\includegraphics[trim=1.5cm 2.9cm 1.5cm 1.5cm, clip=true,width=0.51\textwidth]{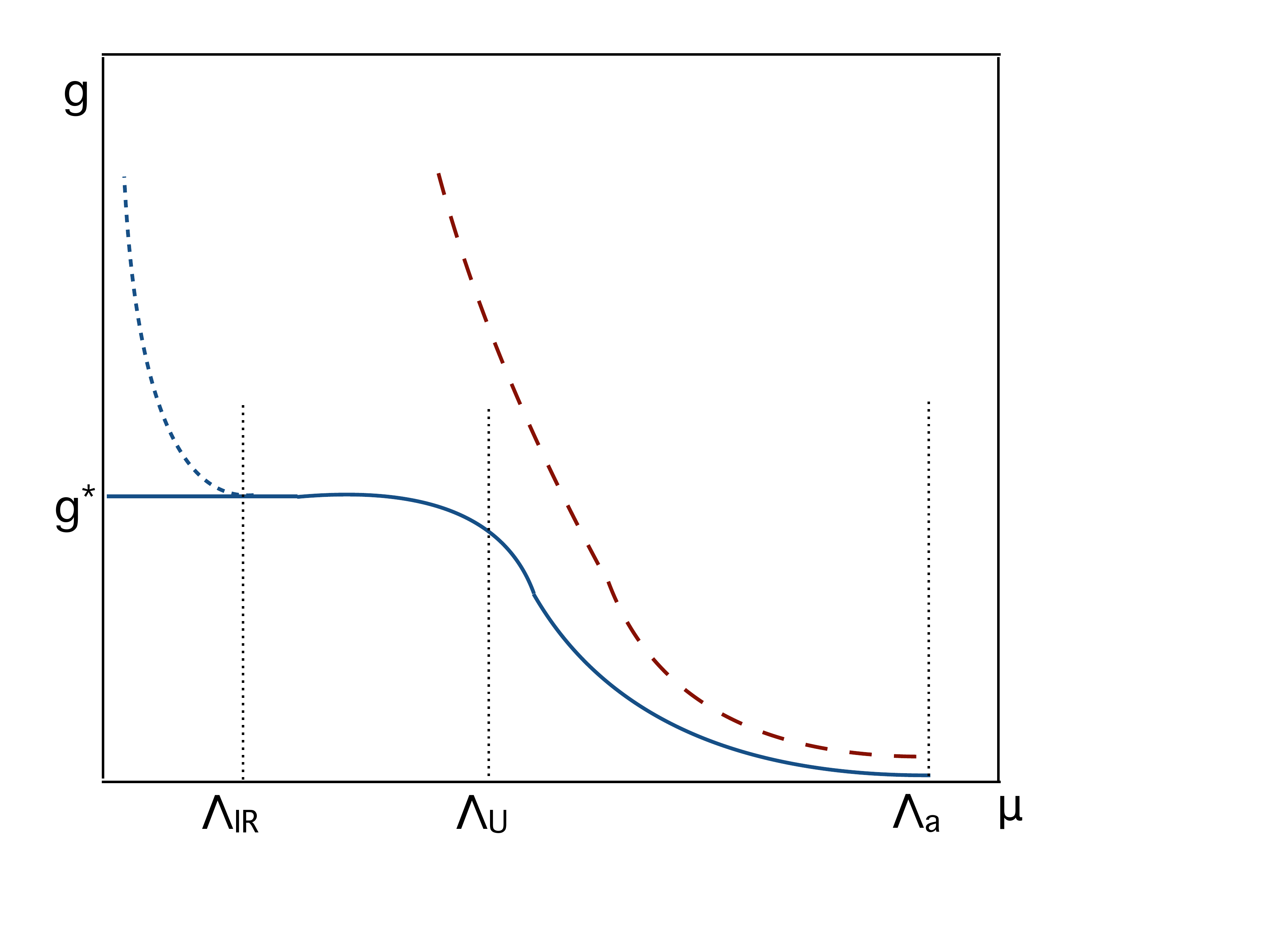}}
\parbox{0.49\textwidth}{\includegraphics[width=0.47\textwidth]{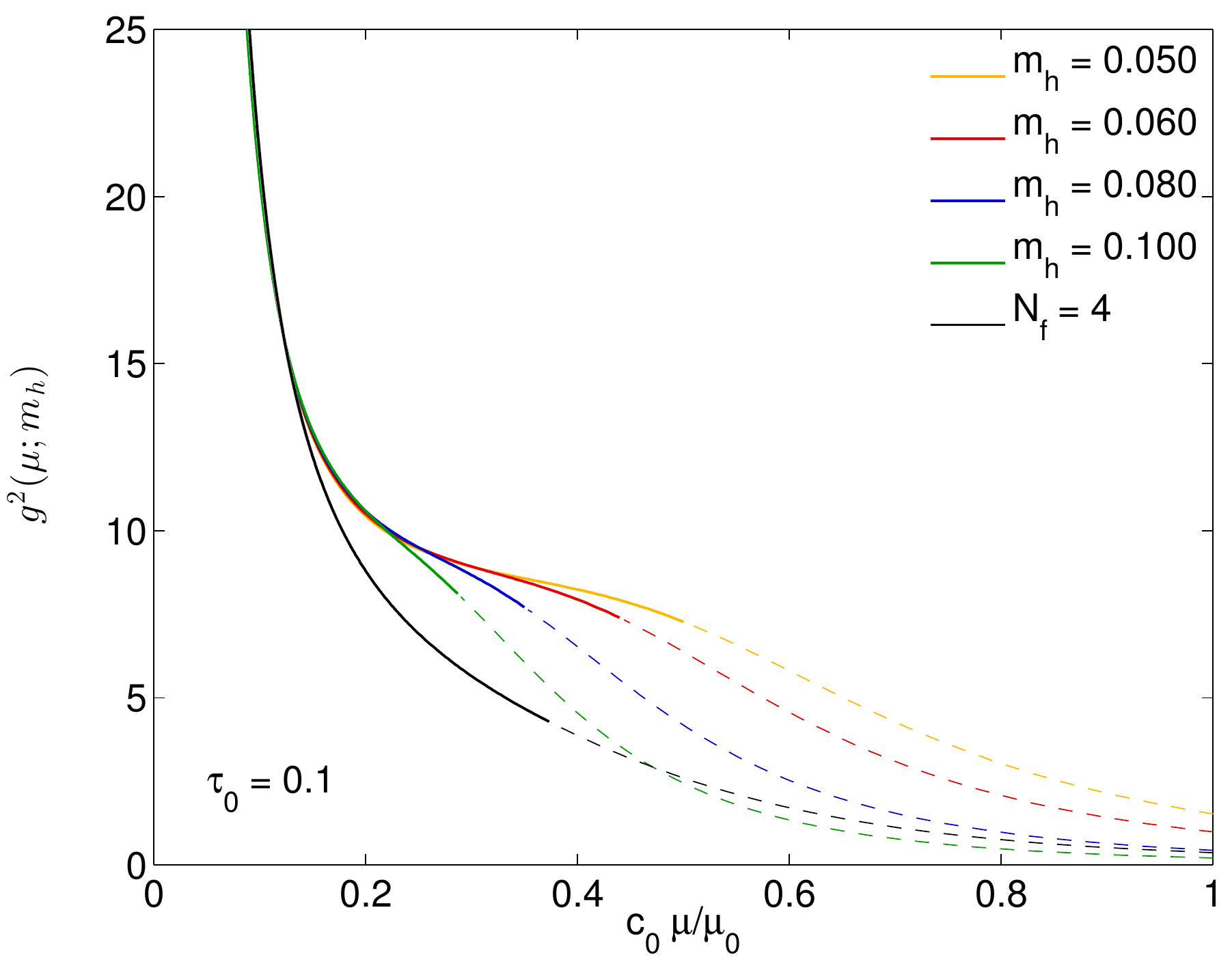}}
\caption{Left: The expected running gauge coupling of conformal and mass-deformed system. The solid blue curve sketches the evolution of the gauge coupling with the energy scale in a conformal system. The dashed blue curve shows the modification in a mass deformed system while the red long-dashed curve corresponds to a situation where the fermions decouple before the gauge coupling could approach the conformal IRFP. Right:  The running coupling constant $\gtGF$ at the mass scale $\mu$ for different values of $m_h$  
with $m_\ell$ extrapolated to the chiral limit.  $\mu_0$ and $c_0 = \mu_0^{-1}|_{m_h=0.050}$ serve as normalization constants that ensure that the different systems are compared at matching energy scales  and $\tau_0$ is the shift parameter to remove discretization errors. The dashed sections of the lines indicate where we  suspect cutoff effects  may be significant. The running gauge coupling in our model at four different $m_h$ mass values and in the $m_h=\infty$ 4-flavor limit. The emergence of the walking regime is evident as $m_h \to 0$. }
\label{fig:walking_coupling}
\end{figure}

\paragraph{a) Conformal systems} 
The solid curve in \fref{fig:walking_coupling} depicts  the running gauge coupling of a conformal system.  All fermion masses are $\widehat m_i = 0$ and the quantity  $\Lambda_a=\pi/a$ denotes the lattice scale where the bare parameter $g_0$ is set. If we  want to define a continuum theory at the perturbative UVFP the bare gauge coupling $g(\Lambda_a)$ has to be tuned to $g(\Lambda_a) \to 0$. As we lower the energy scale from $\Lambda_a$ towards the IR, the gauge coupling  changes according to \eref{eq:rg_running},  i.e. it  approaches the IRFP.   We denote the scale where  the gauge coupling gets close enough to the  IRFP  such that $g(\mu)-g^\star$ can be neglected by $\Lambda_U$, the UV scale where the dynamics governed by the IRFP sets in.  The running gauge coupling at this scale takes the value of the IRFP,     $g(\mu) \approx g^\star$ and approaches $g^\star$ as $\mu \to 0$.

\paragraph{b) Mass deformed system}
Next  we consider the case where all fermions acquire a small bare mass at  the cutoff scale $\Lambda_a$.  The running of the gauge coupling is only minimally affected until $\widehat m_i(\mu) = \mathcal{O}(1)$. We denote the scale where $\widehat m_i(\mu) = 1$ as $\Lambda_{IR}$, the IR scale where the fermions decouple.  As the system runs deeper in the IR, the  gauge coupling follows the running coupling of a pure gauge system. This behavior is denoted by the blue dashed curve in \fref{fig:walking_coupling}. Unless the bare fermion mass is tuned $\widehat m(\Lambda_a) \to 0$, in the deep IR a mass deformed system has only gauge degrees of freedom. That does not mean, however, that the system is equivalent to a pure gauge system, as we have discussed already in section \ref{sec:rgequations}.

\paragraph{c) Mass deformed systems far from the IRFP}
If the fermion mass at the cutoff level is too large, it is possible that  $\Lambda_{IR} > \Lambda_U$. In that case the fermions decouple before the gauge coupling reaches the IRFP. This scenario is depicted by the red long-dashed lines in  \fref{fig:walking_coupling}. In this case the IR dynamics is close to the pure gauge system with heavy fermions,  and is not of interest for us.

\paragraph{d) Mass deformed system with split fermion masses}
In our final case we keep $N_\ell$  fermions massless  and assign finite, degenerate bare mass $\widehat m(\Lambda_a)$ to the others. We will assume that $\widehat m(\Lambda_a)$ is small enough that $\Lambda_{IR} \ll \Lambda_U$ and that in the deep IR, where only the $N_\ell$ massless fermions contribute to the dynamics the system is chirally broken, i.e.~$N_\ell$ is below the conformal window. Qualitatively the running coupling is very similar to  that of the mass deformed system as discussed in case {\it b)}. While the non-zero fermion masses are small the gauge coupling behaves like in a conformal system. At the $\Lambda_{IR}$ scale the massive fermions decouple, chiral symmetry breaks and the IR system has $N_\ell$ massless fermions. The running gauge coupling is driven by this dynamics, similarly to the dashed blue line in  \fref{fig:walking_coupling}. Just like in case {\it b)} this does not mean that the deep IR dynamics is equivalent to an SU($N_c$) gauge system with $N_\ell$ massless fermions. Before decoupling, the additional fermions have influenced the IR dynamics, leading to specific hyper scaling relations as we discussed in section \ref{sec:rgequations}.

\subsection{The running coupling of the 4+8 flavor system}

The gradient flow (GF) transformation provides a natural definition of the running coupling that is easy to implement numerically
\cite{Narayanan:2006rf,Luscher:2009eq,Luscher:2010iy}.
The GF running coupling at energy scale $\mu$ is defined as
 \begin{align}
g^2_{GF}(\mu) = \frac{1}{\mathcal{N}} \langle t^2 E(t) \rangle,
\label{eqn8}
\end{align}
where $t=a^2 t_{\rm{lat}}$  ($t_{\rm{lat}}\gg 1$) is the flow time that is related to the energy scale as  $\mu^{-1} = \sqrt{8t}$. The energy density 
\begin{align}
E(t) = -\frac{1}{2}{\rm Re Tr}[G_{\mu\nu}(t)G^{\mu\nu}(t)]
\label{eq7}
\end{align}
 can be evaluated by any appropriate lattice operator while the  constant $\mathcal{N}= 3(N^2-1) / 128\pi^2$ is chosen such that $g^2_{GF}$ matches the traditional $\overline{MS}$ coupling in perturbation theory \cite{Luscher:2010iy}.

 We  show $\gGF(\mu)$ as  function of the energy $\mu$ from our numerical simulation of the 4+8 flavor system in the right panel of \fref{fig:walking_coupling}. We consider  four  different  values  for the  heavy flavors and extrapolate to the chiral limit in the light quark masses. The right plot in \fref{fig:walking_coupling} also shows the running coupling of the 4-flavor ($m_h=\infty$) system. We rescale $\mu$ by the lattice scale $\mu_0^{-1}=\sqrt{8{t_0}}$ (see~\cite{Luscher:2010iy}) in each case such that the running couplings match in the IR. The dashed curves correspond to flow time where $\sqrt{8t} \le 2.0$ and  cutoff effects could   be important. While the choice  $\sqrt{8t} = 2.0$  to separate discretization effects from physical behavior is rather arbitrary, the trend as the mass of the 8 heavy flavors decrease from $\infty$ towards the chiral limit is clear. The resemblance between the RG inspired behavior depicted  on the left panel and the numerically observed prediction of the right panel is striking though not unexpected: both show the theoretically well motivated running coupling of mass deformed systems near the conformal IRFP.

\section{Numerical Simulations}
\begin{figure}[tb]
\centering
\includegraphics[width=0.49\textwidth]{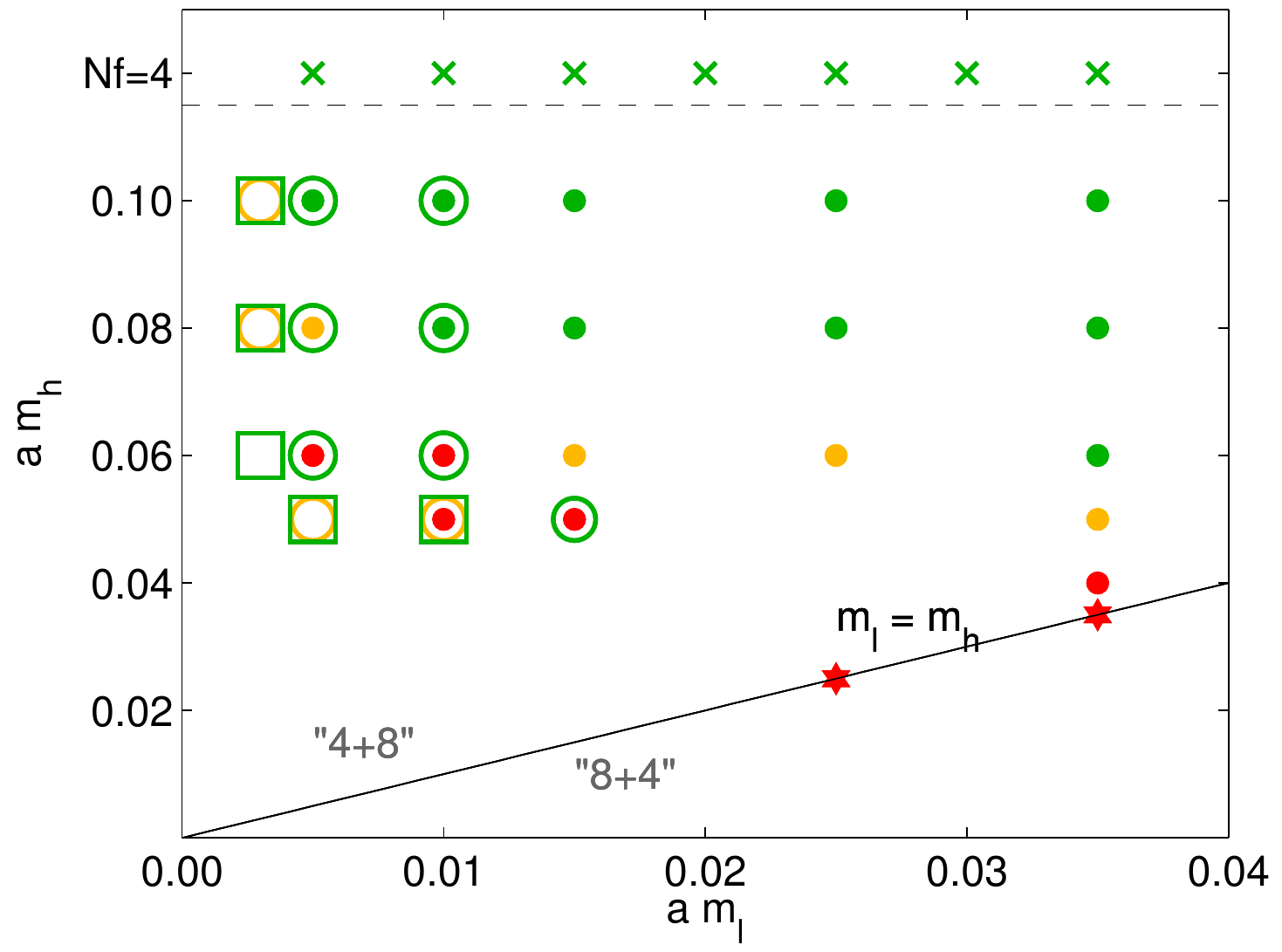}
 \caption{Light ($m_\ell$) and heavy ($m_h$) mass values for the simulations carried out on $24^3\times 48$ lattices (filled symbols) and $32^3\times 64$ lattice (open circles). The colors are meant to caution about finite size effects, likely negligible for green, but of increasing importance as the color turns to orange and red. }
    \label{fig:overview}
\end{figure}

Numerical simulations of our model with four light and eight heavy flavors are carried out using nHYP-smeared \cite{Hasenfratz:2007rf}  staggered fermions and the plaquette gauge action with fundamental and adjoint terms \cite{Cheng:2013xha,Cheng:2013bca}. In previous works\cite{Cheng:2013xha,Cheng:2013bca} with eight and twelve fundamental flavors, these actions  demonstrated small taste breaking artifacts 
and with our choice of smearing parameters the numerical simulations are stable  everywhere. 
We generated a large set of gauge field configurations with different bare input masses for the light and the heavy fermions at a single gauge coupling, $\beta=4.0$.  This gauge coupling is close enough to the $N_f=12$ IRFP that the simulations can be considered to be in the mass deformed region. Starting with small volumes, $L^3\times T=24^3\times 48$, we eventually increased the lattice sizes to  $32^3\times 64$, $36^3\times 64$, and even $48^3\times 96$  to study and account for finite volume effects. We present a graphical overview of our set of ensembles in \fref{fig:overview} and indicate by the color the significance of finite volume effects: green (likely negligible), orange (moderate), red (severe).  All gauge fields were generated by the hybrid Monte Carlo (HMC) update algorithm \cite{Duane:1987de} using the implementation in the FUEL software package \cite{FUEL}.

As usual we monitor algorithmic parameters and physical observables during the evolution of gauge field configurations maintaining e.g.~an average acceptance rate of greater 65\% on all ensembles. Likewise we measure fermionic ($\langle \psi \bar\psi\rangle$) and gluonic (average plaquette) quantities and their dependence on the four space-time coordinates to detect artificial lattice phases. So far all our simulations are consistent with chiral symmetry breaking and do not exhibit any signs of the S4-broken lattice phase \cite{Cheng:2011ic}.

\section{Spectrum}

We have investigated the light fermion spectrum and  the dependence of ``hadronic'' masses on the heavy input quark mass $m_h$.  As we discussed in  \sref{sec:rgequations} we expect hyperscaling in the heavy mass $m_h$ when the light mass $m_\ell$ is extrapolated to the chiral limit. Consequently dimensionless ratios should show independence on $m_h$ up to lattice corrections.

In analogy to QCD we investigate ``mesonic'' quark-antiquark pairs, studying in particular the pseudoscalar (pion, $\pi$), vector (rho, $\varrho$), axial-vector ($a_1$), and isotriplet scalar ($a_0$) bound states. Moreover, we determine the lowest mass baryon (three quark bound state), the nucleon ($n$). These quantities are described by quark-line connected diagrams, thus the numerical determination is straightforward and we obtain our results using propagators with wall-sources. In addition we determine the mass of the  isosinglet scalar ($0^{++}$) meson bound state which has the same quantum numbers as the Higgs boson. Determining the isosinglet scalar is much more challenging because quark-line disconnected diagrams also contribute. These disconnected diagrams are numerically much noisier and require different numerical methods.

\subsection{Connected light fermion spectrum}
\begin{figure}[tb]
\centering
\parbox{0.49\textwidth}{\includegraphics[width=0.49\textwidth]{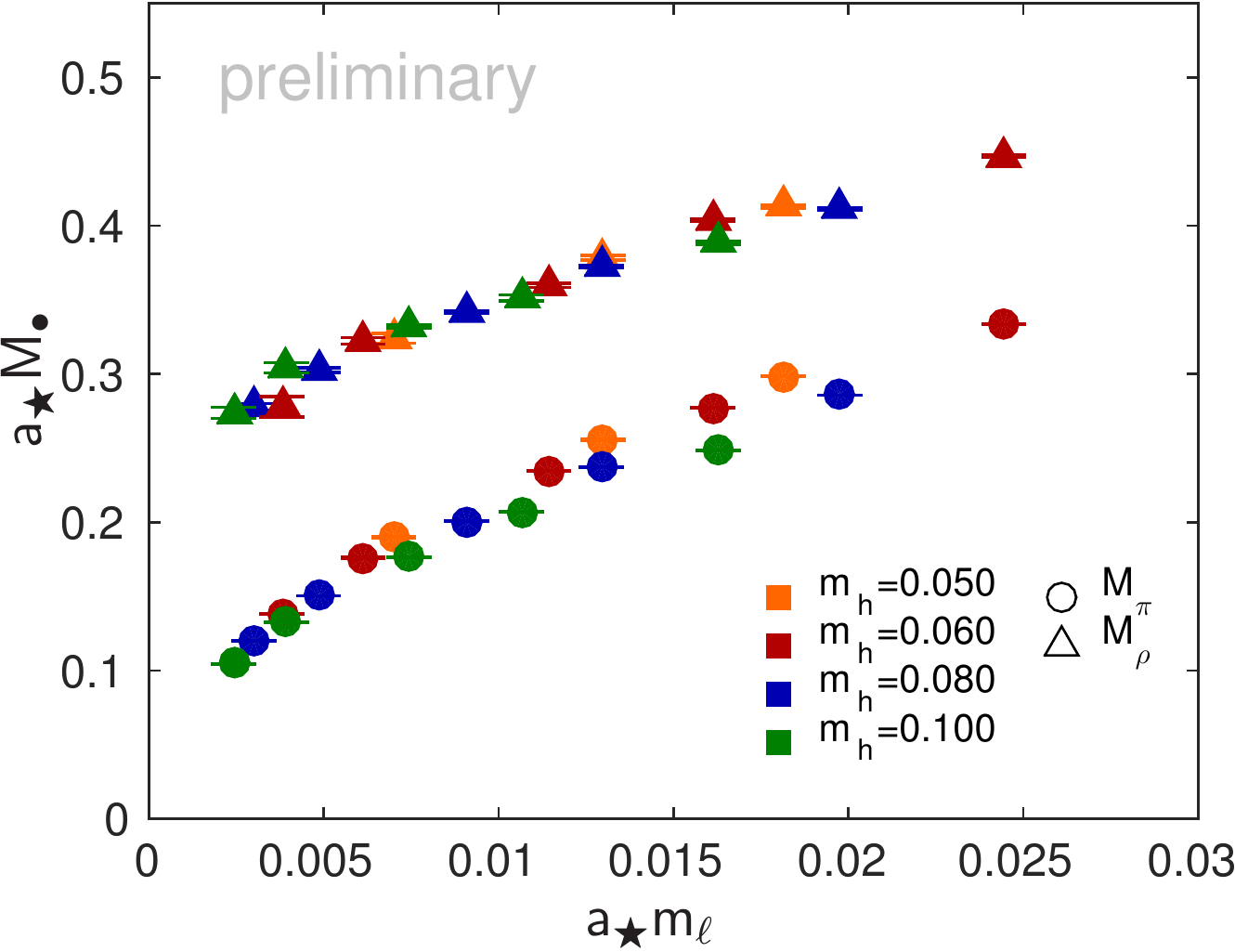}}
\parbox{0.49\textwidth}{\includegraphics[width=0.49\textwidth]{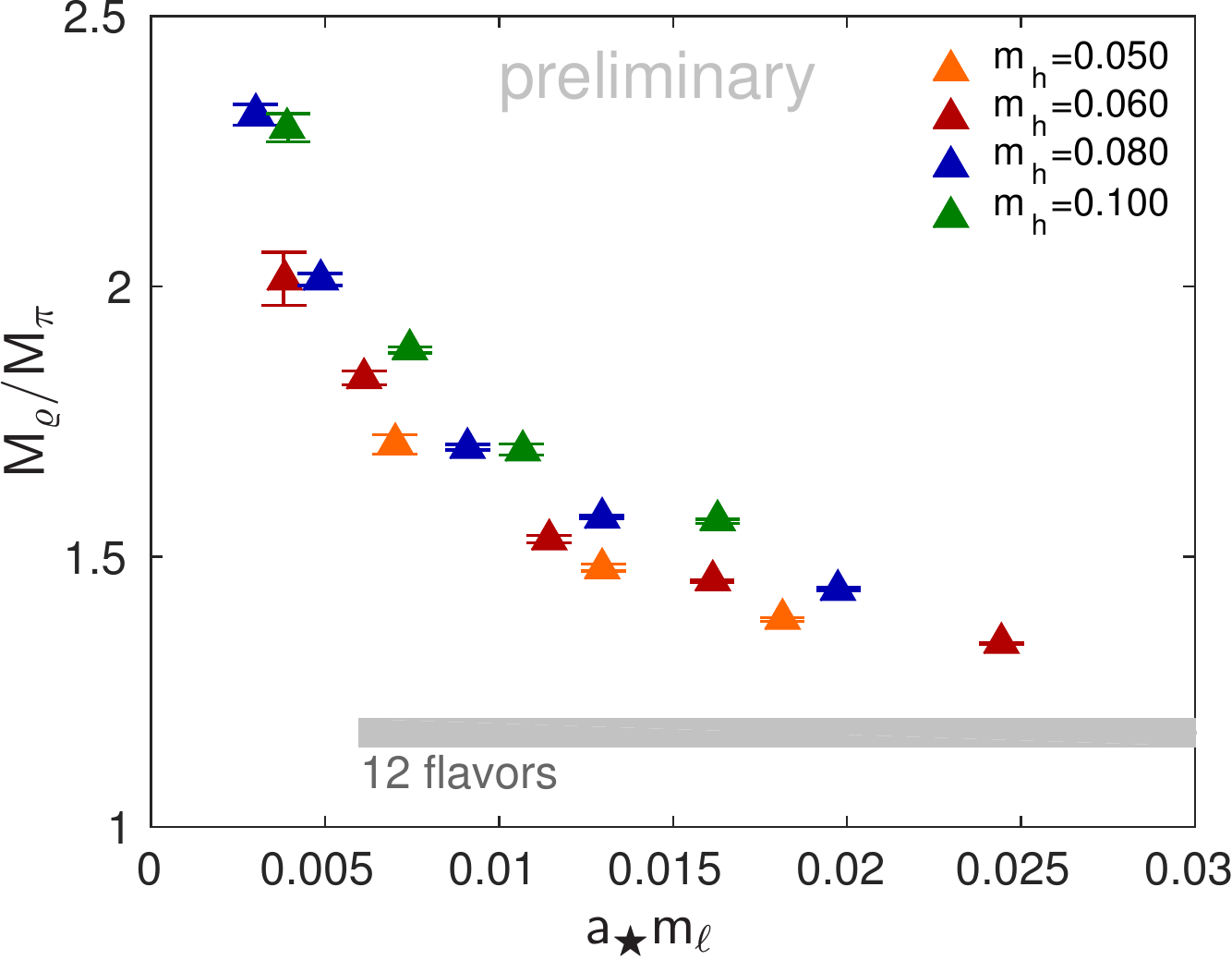}}
\caption{Left: Pion and rho vs.~light input quark mass shown in lattice units $a_\bigstar$ defined  on our  $36^3\times 64$, $m_h=0.080$, $m_\ell=0.003$ reference ensemble. The rho shows (except for small finite volume effects) no dependence on $m_h$ and the pion exhibits only a very weak dependence, most pronounced at larger $m_\ell$ values. Right: plotting the ratio $M_\varrho / M_\pi$ vs.~the light input quark mass confirms that for $m_\ell\to 0$ chiral symmetry is broken for all our choices of $m_h$. As reference the value of $M_\varrho/M_\pi$ found in the corresponding 12 flavor system~\cite{Cheng:2013xha} is indicated by the grey band.}
\label{fig:ConnSpectrum}
\end{figure}

Using ensembles with sufficient number of thermalized trajectories and sufficiently large volumes, the left panel of ~\fref{fig:ConnSpectrum} shows our results for the pion and the rho. We present the masses in units of a lattice reference scale $a_\bigstar$ which we  define  through the gradient flow $\sqrt{8t_0}$ scale on our $36^3\times 64$, $m_h=0.080$, $m_\ell=0.003$ reference ensemble. Values in lattice units obtained on other ensembles are converted by multiplying appropriate ratios of the gradient flow scale \cite{Narayanan:2006rf,Luscher:2010iy} which we determine on all ensembles.

As one expects based on the arguments in  \sref{sec:rgequations}, the dependence 
      on the input quark mass  $m_h$ is weak. Except for small finite volume effects, our measurements of the rho mass shows no dependence on $m_h$. The pion mass, $M_\pi$ shows the same overall trend; however a small dependence on $m_h$,  in particular  for the heavier $m_\ell$ data sets is visible. Hyperscaling in $m_h$ is only expected in the $m_\ell = 0$ limit. Our data are consistent with this expectation.
      
       In the right panel of \fref{fig:ConnSpectrum} we show the ratio $M_\varrho/M_\pi$ as function of the light quark mass. In contrast to a mass deformed conformal system (e.g.~12 flavors \cite{Cheng:2013xha}), where all hadrons scale the same way with respect to the mass, we observe that for all our choices of $m_h$ the ratio  $M_\varrho/M_\pi$ is consistent with a divergent behavior  for $m_\ell \to 0$. We take this as a certain indicator that in our simulations chiral symmetry is  broken spontaneously in the limit of vanishing light quark mass.   

In addition to the hadronic masses we also determine the pseudoscalar decay constant $F_\pi$ which is the preferred choice for setting the scale in relation to EWSB.   The quantity $F_\pi L$  controlling the chiral perturbative expansion  is $1.0 \lesssim F_\pi L \lesssim 1.35$ for all our relevant data sets. However the smallness of this parameter is not the only problem in performing a chiral extrapolation. The  $0^{++}$ state, as will be discussed in the next section,  is nearly degenerate with the pion,  rendering standard chiral perturbative expansions and consequently chiral extrapolations questionable.

\subsection{Determination of the isosinglet $0^{++}$ scalar}
The determination of the isosinglet $0^{++}$ meson is more complicated and requires the evaluation of disconnected diagrams. We construct the disconnected operator from the vacuum subtracted operator
\begin{align}
{\cal O}_\text{disc} = \langle \bar \psi \psi\rangle(t) - \langle \langle\bar \psi \psi\rangle\rangle_e,
\label{eq:O_disc}
\end{align}
and determine the vacuum contribution via the ensemble average $\langle \langle\bar \psi \psi\rangle\rangle_e$.  The operator $\langle \bar \psi \psi\rangle(t)$ is measured using $N_r=6$ full volume noise sources  diluted in time, color, as well as spatially even/odd in order to reduce the stochastic noise.\cite{Foley:2005ac}

In systems with degenerate  flavors the $0^{++}$ state is the ground state of the correlator $N D(t) - C(t)$ where $D(t)$ and $C(t)$ are the disconnected and connected correlators of the scalar operator $\langle \bar \psi \psi\rangle(t)\,\langle \bar \psi \psi\rangle(0)$ and $N$ denotes the number of flavors. When using staggered fermions $N=N_f/4$. In our case the physical $0^{++}$ correlator is more complicated as it is a  mixture of  light-light, light-heavy, and heavy-heavy scalar states. In addition the connected correlator couples strongly to excited states, making the determination of the $0^{++}$ numerically challenging. However,  as  was pointed out in Ref.~\cite{Aoki:2013zsa}, the ground state mass can be determined from the disconnected correlator alone. $D(t)$ couples to the $0^{++}$ state, and since that state is the lightest, it dominates the asymptotic behavior. Using only the disconnected correlator  provides a numerically more stable determination of the lowest isosinglet scalar mass and avoids the cumbersome mixing of the light and heavy states as well. The vacuum subtraction in \eref{eq:O_disc} is the source of large statistical fluctuations.  In practice we find it is better to  fit the finite difference $D(t+1)-D(t)$  of the disconnected correlator  and avoid the vacuum subtraction altogether.  

We perform measurements every 20 MDTU and have extended almost all of our Monte Carlo evolutions to at least 20,000 MDTU, in some cases even twice as much. Nevertheless the determination of the mass from the correlators is particularly difficult for larger values of $m_\ell$. We therefore restrict the presentation of our $0^{++}$ results to the subset of our ensembles with $m_h=0.080$.  We expect hyperscaling in the $m_\ell=0$ chiral limit, therefore the $0^{++}$ mass at any fixed $m_h$ is indicative of its value everywhere in the scaling region of $m_h$.

\section{Summary}

\begin{figure}[tb]
\centering
\parbox{0.49\textwidth}{\includegraphics[width=0.49\textwidth]{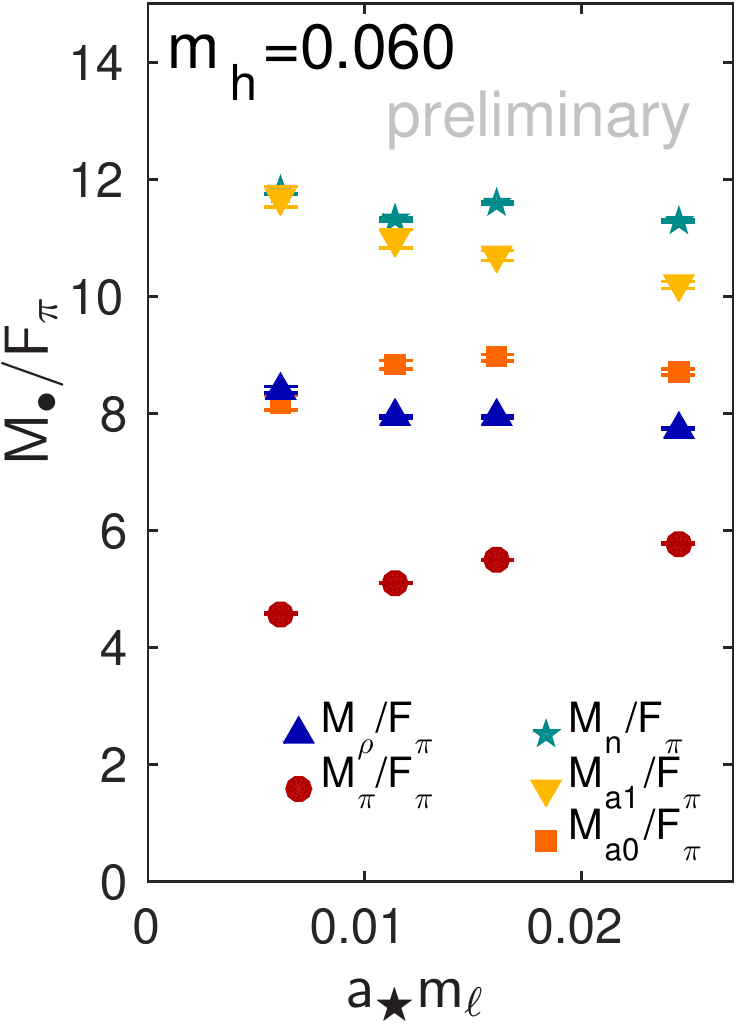}}
\parbox{0.49\textwidth}{\includegraphics[width=0.49\textwidth]{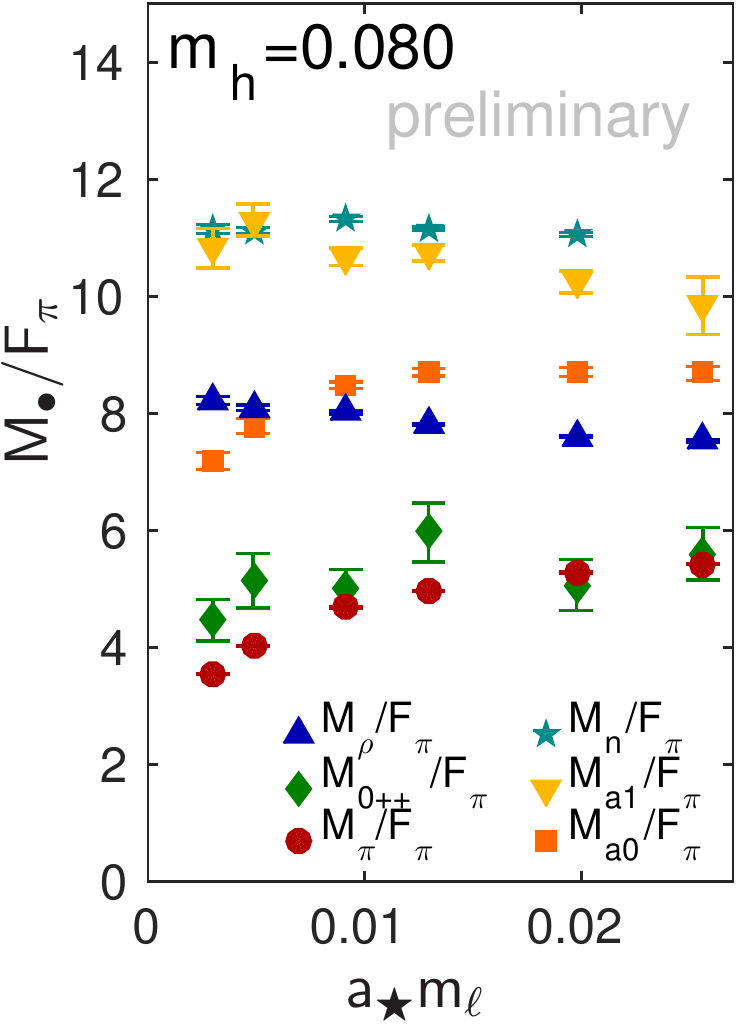}}
\caption{Pion, rho, scalar $0^{++}$, axial, $a_0$, and nucleon mass of the light flavor spectrum in units of $F_\pi$ as function of the light quark mass $a_\bigstar m_\ell$ for $m_h = 0.060$ and $0.080$.  If a  chirally broken system   triggered EWSB,   $F_\pi \approx  250$ GeV would set the correct electroweak  scale. }
\label{fig:Spectrum_080}
\end{figure}

The right panel of \fref{fig:Spectrum_080}  summarizes our results for the $0^{++}$ isosinglet and several other isomultiplet states, normalized by the  pion decay constant $F_\pi$, as the function of the light fermion mass $m_\ell$ at $m_h=0.080$. The left panel shows the isomultiplet states at $m_h=0.060$ where our statistics is not yet sufficient to determine the mass of the $0^{++}$ scalar. Hyperscaling, as we argued in~\sref{sec:rgequations}, implies that all dimensionless ratios are identical, up to lattice corrections, in the $m_\ell=0$ chiral limit. Comparing the two panels of \fref{fig:Spectrum_080} shows that this expectation is satisfied by our data for the connected spectrum. There is no reason to believe that the $0^{++}$ scalar state would be any different. 

 Unlike in QCD, we observe a $0^{++}$ state at $m_h=0.080$ that is nearly  degenerate with the pion and much lighter than the rho. We have not attempted any kind of chiral extrapolation ---   standard chiral perturbation theory is based on the pion being the only light particle in the spectrum which is clearly not true in our case.  Moreover we observe that $M_\varrho/F_\pi$, $M_{a_1}/F_\pi$, and $M_n/F_\pi$  have a fairly linear dependence on $m_\ell$.  
In particular  the ratio $M_\varrho/F_\pi$ appears nearly independent of both $m_\ell$  and  $m_h$ with a value around 8.  Coincidentially  if  $F_\pi\approx 250$ GeV,   the mass of the vector resonance in our model is around 2 TeV, matching the recently reported signal by ATLAS and CMS collaborations\cite{Aad:2015owa, CMS:2015gla}. It is  a curiosity to note that  $M_\varrho/F_\pi \approx 8 $ appears to be a rather general feature of SU(3) gauge fermion systems: this ratio is $ 8.4$  in QCD\cite{Agashe:2014kda};  approximately  8.0 in  the conformal  SU(3) $N_f=12$ model \cite{Fodor:2011tu,Aoki:2012eq}; about $7.9$ with 8 fundamental flavors~\cite{Aoki:2013xza,LSD2015}; approximately 8 in $N_f=2$ flavor SU(3) sextet model~\cite{Kuti2015}. Various theoretical models and approaches also predict that this ratio is independent of the specifics of the IR dynamics, see for example Refs.~\cite{Pagels:1979hd,Appelquist:1986tr}. 

While we expect the dimensionless ratios to be independent of $m_h$, these ratios can depend on the details of the system, like the IRFP of the underlying conformal massless model and the number of  light vs.~heavy flavors.
Comparing different systems could reveal which, if any, of these models are potentially compatible with experimental observations. 

Our work is just the first step in this direction. Choosing a system with two (or four) massless and many massive fermions ensure that in the IR the model is chirally broken while  choosing the total number of fermions such that in the massless limit the system is conformal ensures hyper scaling. While a  similar construction is possible even if the underlying mass-degenerate system is not conformal, it is not obvious what kind of scaling relations one can obtain in that case.

In the future we would like to extend these investigation to systems with 2 massless fermions with $N_f=N_\ell+N_h$ closer to the conformal window. Numerical simulations indicate that systems with $N_f=8$  or 10 fundamental fermions exhibit large mass anomalous dimensions, making them better candidates to satisfy electroweak constraints.

\section*{Acknowledgments}
Computations for this work were carried out in part on facilities of the USQCD Collaboration, which are funded by the Office of Science of the U.S.~Department of Energy, on computers at the MGHPCC, in part funded by the National Science Foundation, and on computers allocated under the NSF Xsede program to the project TG-PHY120002. 
We thank Boston University, Fermilab, the NSF and the U.S.~DOE for providing the facilities essential for the completion of this work. R.C.B., C.R.~and E.W.~were supported by DOE grant DE-SC0010025 and in addition acknowledge the support of NSF grant OCI-0749300. A.H.~acknowledges support by the DOE grant DE-SC0010005. O.W.~is supported by STFC, grant ST/L000458/1. R.C.B., A.H. and C.R.~thank the Aspen Center for Physics, which is supported by National Science Foundation grant PHY-1066293. R.C.B., A.H., and~O.W.~thank the KITP, Santa Barbara, supported in part by the National Science Foundation under Grant No. NSF PHY11-25915.
\bibliographystyle{ws-procs975x65}
\bibliography{../General/BSM}
\end{document}